\documentclass[journal,10pt]{IEEEtran}
\usepackage{amsthm}
\usepackage{amsmath}
\usepackage{graphicx}
\usepackage{algorithm}
\usepackage{algorithmicx}
\usepackage{algpseudocode}
\usepackage{subfigure}
\usepackage{overpic}
\usepackage{colortbl}

\ifCLASSINFOpdf
\else
\fi
\begin{document}
%
\title{ Decompose X-ray Images for Bone and Soft Tissue}
%
%
%

\author{Yuanhao~Gong
	\thanks{Manuscript received April 19, 2005; revised September 17, 2014.}
	}

%
%

\markboth{Journal of \LaTeX\ Class Files,~Vol.~14, No.~8, August~2015}%
{Yuanhao: Asymmetric Kernel}
%



\maketitle

\begin{abstract}
Bones are always wrapped by soft tissues. As a result, bones in their X-ray images are obscured and become unclear. In this paper, we tackle this problem and propose a novel task to virtually decompose the soft tissue and bone by image processing algorithms. This task is fundamentally different from segmentation because the decomposed images share the same imaging domain. Our decomposition task is also fundamentally different from the conventional image enhancement. We propose a new mathematical model for such decomposition. Our model is ill-posed and thus it requires some priors. With proper assumptions, our model can be solved by solving a standard Laplace equation. The resulting bone image is theoretically guaranteed to have better contrast than the original input image. Therefore, the details of bones get enhanced and become clearer. Several numerical experiments confirm the effective and efficiency of our method. Our approach is important for clinical diagnosis, surgery planning, recognition, deep learning, etc.   
\end{abstract}


%
\IEEEpeerreviewmaketitle

%
%
%
%
\section{Introduction}
\label{sec:intro}
\IEEEPARstart{X}-ray has being frequently used in biomedical imaging and clinical diagnosis, especially for bone research and human body diagnosis. X-ray has been studied and developed since 1895. Nowadays, it has become a popular way for bone diagnosis in clinical applications.

Bones are very common in mammals. The bone skeleton provides the basic structure support for mammals such that the body can keep the rigidity when it changes its status (walking, running, or dancing). Bones are also important for the mammals' health. This is one reason that bone research is an important topic for human.

Meanwhile, bones in mammals are usually wrapped by various soft tissues. Such soft tissue helps the growth and development of bones. It also provides some protection for the bones, reducing the possible external force and pressure. 

These bones and their surrounding soft tissues might have various thickness and density. Such property can be used to distinguish them from each other, especially in their images.  

In the X-ray imaging, X-rays are absorbed and scattered by the soft tissue. X-rays are also (significantly) reduced by the dense bones. When the X-rays finally reach the sensors, the images for bones and soft tissues show very different features. 

Dense structure such as bones blocks more X-ray and its image region on the sensor receives less X-ray. Therefore, it is darker than other regions. In contrast, less dense and thin soft tissue does not block much X-ray and its image region on the sensor receives more X-ray. 

To better visualize the bone region, modern X-ray images usually take the residual between a constant maximum value (caused by the given dose) and the original X-ray image, leading to a brighter bone region and a darker soft tissue region.

\begin{figure}[!htb]
	\centering
	\subfigure[Input X-ray Image]{\includegraphics[width=0.3\linewidth]{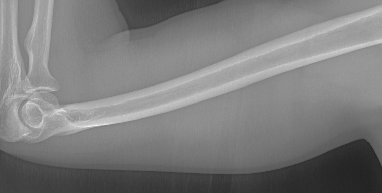}}
	\begin{minipage}{0.05\linewidth}
		\centering
		\vspace{-1.1cm}
		$\Longrightarrow $
	\end{minipage}
	\subfigure[Output Soft Tissue]{\includegraphics[width=0.3\linewidth]{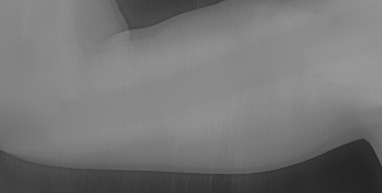}}
	\subfigure[Output Bones]{\includegraphics[width=0.3\linewidth]{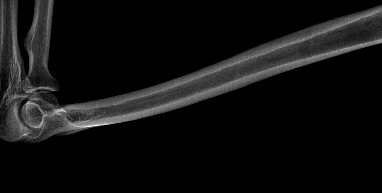}}
	\subfigure[Input X-ray Image]{\includegraphics[width=0.3\linewidth]{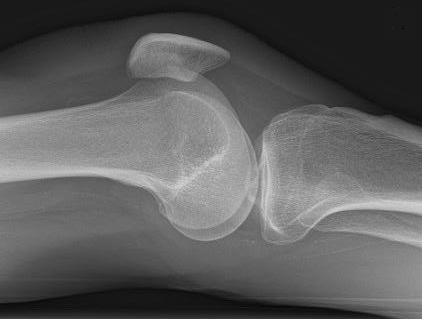}}
	\begin{minipage}{0.05\linewidth}
		\centering
		\vspace{-2.cm}
		$\Longrightarrow $
	\end{minipage}
	\subfigure[Output Soft Tissue]{\includegraphics[width=0.3\linewidth]{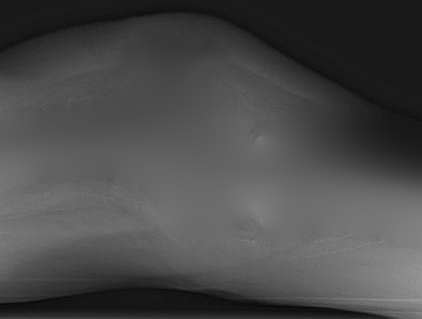}}
	\subfigure[Output Bones]{\includegraphics[width=0.3\linewidth]{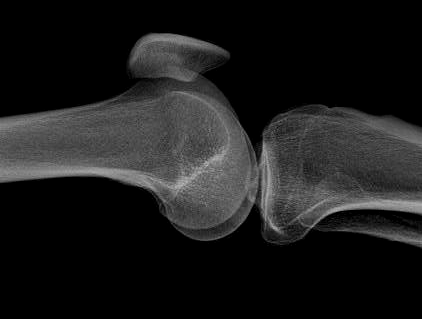}}
	\caption{Original X-ray images (left), estimated soft tissue (middle) and estimated bones (right). The estimated bones are theoretically guaranteed to have better image contrast than the input image. Therefore, the details on bones are enhanced.}
	\label{fig:illu}
\end{figure}

In such X-ray image, bone region has higher image intensity and can be seen clearly. And we can set different intensity range to visualize the bone region (Window Technique). However, such visualized bone regions are still composed by bones and soft tissues because they are overlapped from the sensor point of view. 

To obtain better bone visualization, some other methods try to increase the image contrast, such as histogram equalization, Contrast Limited Adaptive Histogram Equalization (CLAHE), etc. However, they {\bf simultaneously} enhance the bone and soft tissue. Even though the image becomes visually better, the relationship between image intensity and actual X-ray dose becomes complex (even unknown), leading to difficulties for clinical diagnosis.

To tackle these problems, we propose to estimate the soft tissue image and bone image simultaneously without losing the linear relationship between image intensity and physical property of the imaging objects. Two examples from our method is shown in Fig.~\ref{fig:illu}. The bone details are enhanced, which is theoretically guaranteed. The details of our method will be explained in later sections.

\subsection{Scattering Light in Physics}
As shown in the left column of Fig.~\ref{fig:illu}, bones are usually surrounded by the soft tissue. This physical configuration is similar with many natural scenes. One example is the foggy weather, as shown in Fig.~\ref{fig:phy} (a). The fog can be considered as ``soft tissue'' (low density) and the buildings can be considered as ``bone'' (high density). 

The physics behind this phenomena is the scattering light~\cite{Hulst}. Scattering light is common in various scenarios, such as X-ray images in clinics, foggy weather in natural scene, and fluorescence images in biological images~\cite{gong:gdp,gong:phd}. The scatter light might downgrade the image quality. For example, the soft tissue in human body scatters the X-ray, making the bone details unclear. 

The scattering light has been studied long time ago. In 1871, Rayleigh studies this physics when the wavelength of light is larger than the radius of particles in the medium~\cite{Hulst}. The more general case with ball shape particles were studied by Mie and it is called Mie scattering. These physics achievements give the theoretical foundation for modern image dehazing approaches. 

\subsection{Scattering Light in Natural Images}
\begin{figure}[!tb]
	\centering
	\subfigure[natural]{\includegraphics[width=0.35\linewidth]{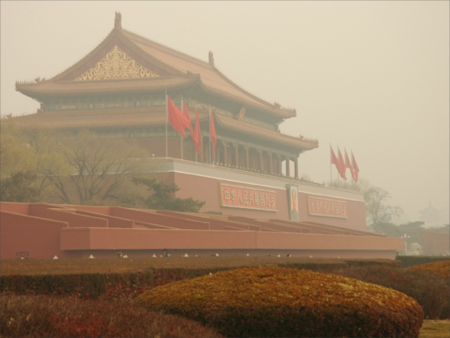}}
	\begin{minipage}{0.2\linewidth}
		\centering
		\vspace{-2cm}
		dehaze\\
		$\Longrightarrow $
	\end{minipage}
	\subfigure[remove fog ]{\includegraphics[width=0.35\linewidth]{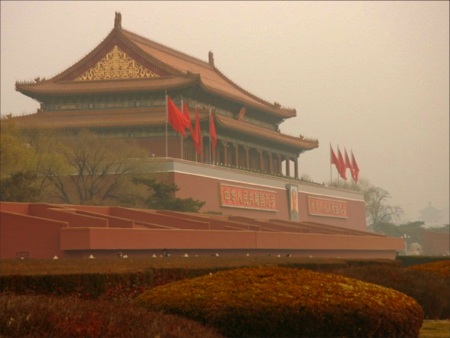}}

	\subfigure[X-ray image]{\includegraphics[width=0.35\linewidth]{images/humerus.png}}
	\begin{minipage}{0.2\linewidth}
		\centering
		\vspace{-1.6cm}
		decompose bone and soft tissue\\
		$\Longrightarrow $
	\end{minipage}
	\subfigure[bone image]{\includegraphics[width=0.35\linewidth]{images/humerus_U.png}}
	\caption{The building is surrounded by the fog (a). The fog can be removed by computation algorithms (b). The bone is surrounded by the soft tissue (c). The soft tissue can be removed by our method (d).} 
	\label{fig:phy}
\end{figure}

A typical example of scattering light in nature is the fog. As a result, the image from foggy natural scene is not clear. Image processing algorithms that virtually remove the fog are called dehazing (as shown in the top row of Fig.~\ref{fig:phy}). 

For natural images, the dehazing mathematical model is simplified as~\cite{fattal:2008,DarkPrior}
\begin{equation}
\label{eq:model}
f(x,y)=J(x,y)t(x,y)+A(1-t(x,y))\,,
\end{equation} where $f$ is the observed image, $J$ is the unknown clear image to be estimated, $t$ is the transmission map to be estimated, and $A$ is the global atmospheric light to be estimated. 

In the past few years, dehazing algorithms have made a significant progress. These methods can be categorized into three types: simple contrast enhancement~\cite{Tan2008,fattal:2008}, dark channel based methods~\cite{DarkPrior} and deep learning methods~\cite{Cai2016}.

Early work considers the foggy images do not have enough contrast and they simply increase the contrast~\cite{Tan2008,fattal:2008}. Such methods are suffering from heavy computation and usually have obvious artifacts in the result. 

One important achievement is the dark channel prior based dehazing methods~\cite{DarkPrior}. Dark channel states that there must be a low value intensity in a local neighborhood region. And the resulting model can be efficiently solved by the guided image filter, which popularizes the dark channel prior.

Deep learning is another type of scattering light removal methods~\cite{Cai2016,Engin2018}. It assumes the clear and foggy image pairs are given and the process that maps the foggy image to its corresponding clear image can be implicitly learned by a neural network. In such methods, the foggy images are usually synthetic. The paired images in practical applications are difficult to be obtained.

\subsection{Bone Suppression}
Instead of interested by bones, some applications focus on the soft tissue such as pneumonia. For these applications, they try to reduce the visualization of bones. Such task is called bone suppression~\cite{Suzuki2006,Chen2014,Li2020}.

Such methods require strong prior information about the imaging objects, such as the rib shapes for Chest X-ray images. And they usually require to exactly find the bone boundaries (bone segmentation). Such methods are difficult to be extended from one imaging object to other imaging objects. For example, the methods developed for rib can not easily be used for feet or knees images.

Even with the accurate bone segmentation, the resulting soft tissue images may have obvious artifacts because their assumptions are not always valid for the input images. 

\begin{figure}[!tb]
	\centering
	{\includegraphics[width=0.6\linewidth]{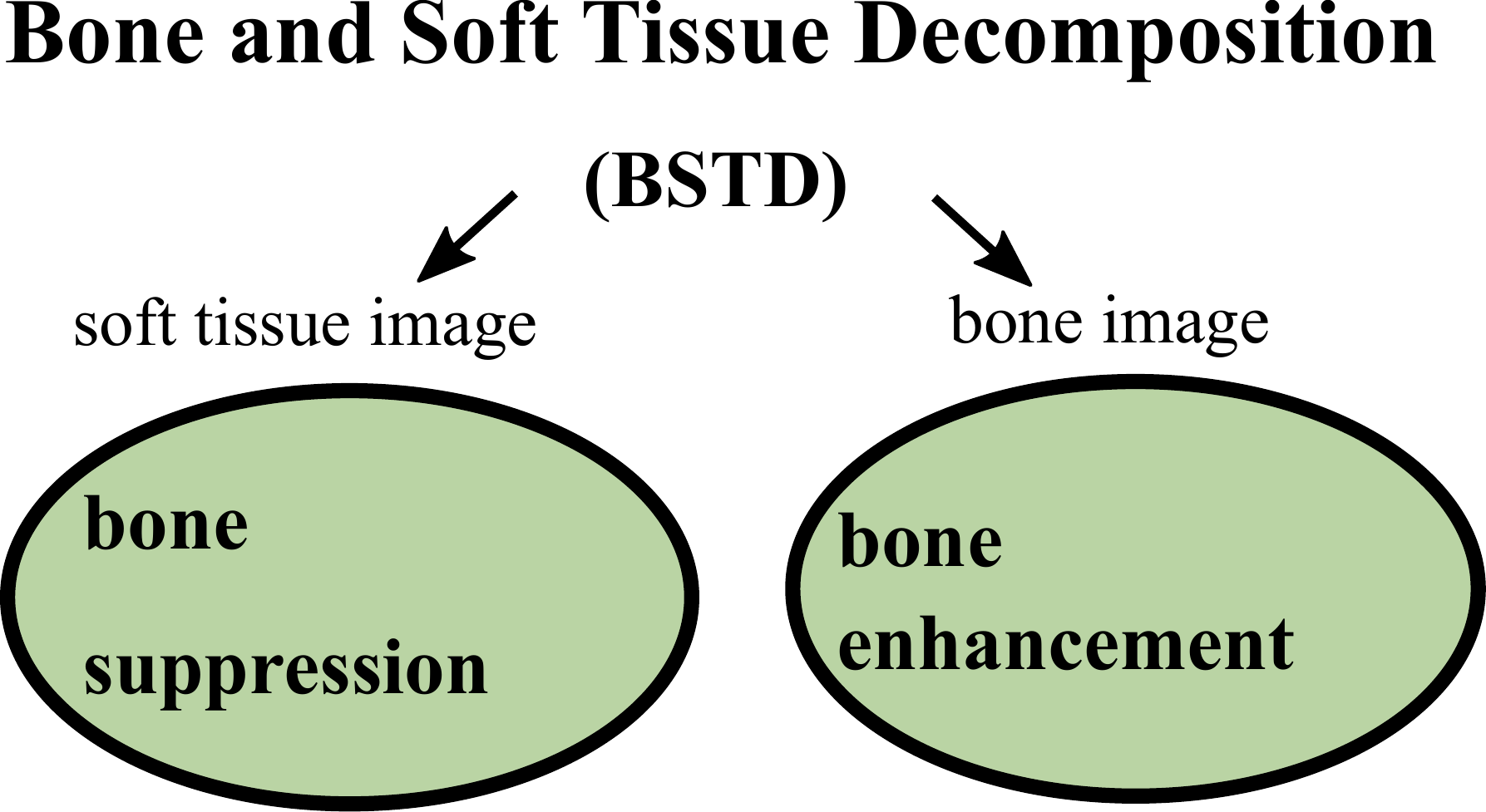}}
	\caption{Our method simultaneously performs bone enhancement and bone suppression.}
	\label{fig:role}
\end{figure}

These limitations motivate us to develop a new and generic mathematical model. Instead of suppressing bones or soft tissue, our model decompose one X-ray image into one soft tissue image and one bone image. These two images have exactly the same imaging domain. Our task is fundamentally different from bone enhancement task and bone suppression task. In fact, our method {\textbf{simultaneously}} does bone enhancement and bone suppression. As illustrated in Fig.~\ref{fig:role}, our soft image can be considered as bone suppression while our bone image can be considered as bone enhancement.
\subsection{Motivation and Contributions}
The soft tissue in human body usually scatters the X-ray, severely reducing the quality of bone details in the resulting images. This fact motivates us to construct a novel mathematical model that can decompose bones and the soft tissue in X-ray images. The decomposed soft tissue image can be used for its related study such as pneumonia. The bone image can be adopted for its related research such as bone fracture.

The scattering light in X-ray images by the soft tissue shares the same physical law as the fog in natural images~\cite{gong:gdp,gong:phd}. Thus, the dehazing methods that have been developed for natural foggy images must be also valid on X-ray images~\cite{Gong2019}. We adopt the transfer learning in machine learning community by applying the dehazing model onto X-ray images. 

\begin{figure}[!htb]
	\centering
	{\includegraphics[width=0.9\linewidth]{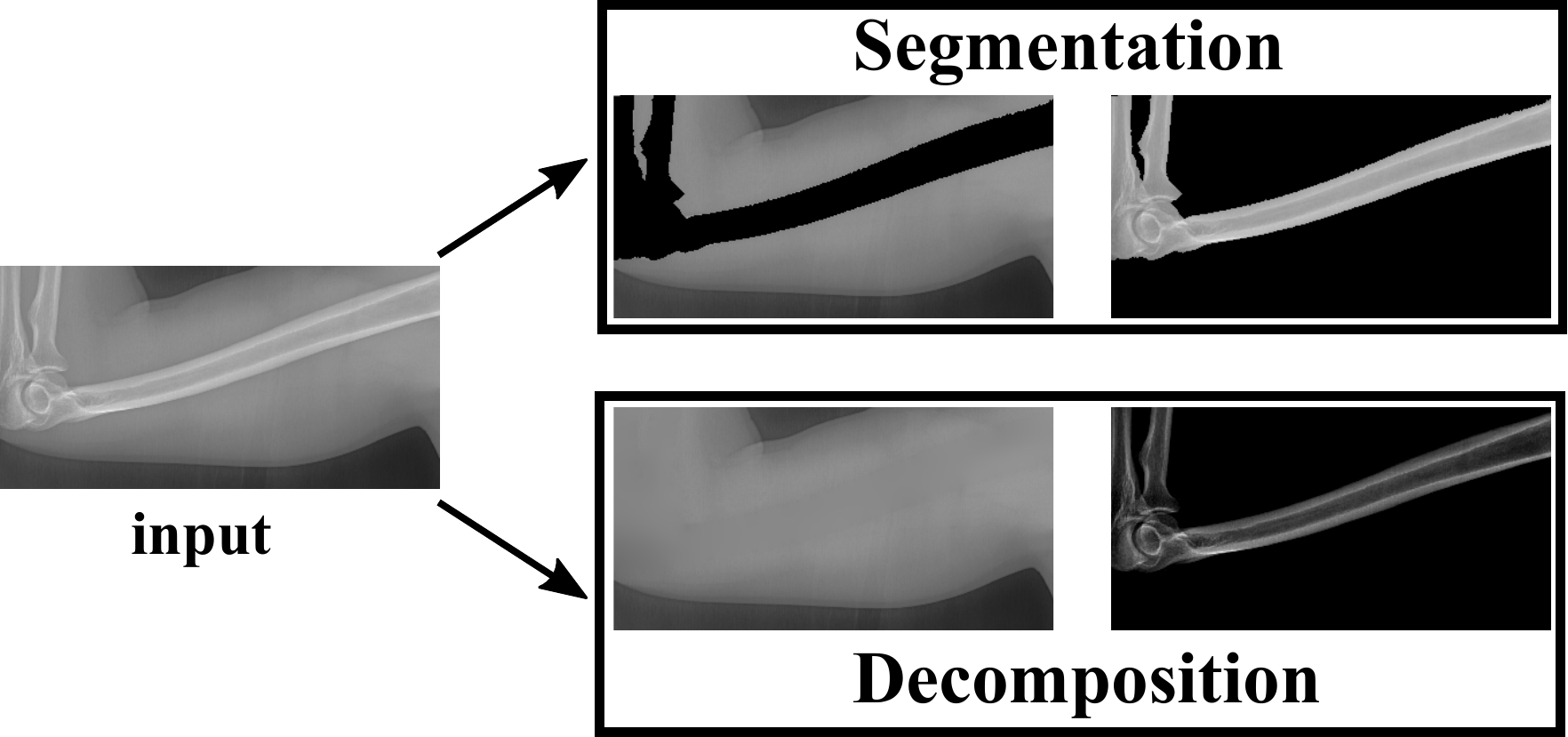}}
	\caption{Our task is different from the classical bone segmentation.}
	\label{fig:seg}
\end{figure}

Different from the bone enhancement or suppression, we propose to decompose the input X-ray image into one bone image and one soft tissue image. Such task is named as Bone and Soft Tissue Decomposition (BSTD). We construct a new mathematical model that can effectively decompose the soft tissues in X-ray images. Our method decomposes the input X-ray image into background image (soft tissue) and bone image. Be aware the difference between our model and the bone segmentation task. Bone segmentation separates the imaging domain into bone region and background region (without overlap). However, our background and bone images share the same imaging domain (exactly overlapped with the same imaging domain). Such difference is illustrated in Fig.~\ref{fig:seg}. 

Our contributions are in following folds:
\begin{itemize}
	\item We propose a new image processing task named as Bone and Soft Tissue Decomposition (BSTD).
	\item We propose a new mathematical model for BSTD. This model is based on the well known image dehazing model, but with proper assumptions for X-ray images.
	\item With some assumptions, the BSTD model leads to a standard Laplace equation, which can be efficiently solved.
	\item The resulting bone image is theoretically guaranteed to have better image contrast than the original input image.
\end{itemize}

\section{Our Method}
In this section, we first show the novel mathematical model that decomposes the soft tissue and the bones. Then, we introduce some proper assumptions to make our model well-posed. With these assumptions, our model leads to a standard Laplace equation, which has an efficient solver. Finally, our model can be efficiently solved. Our method estimates one soft tissue image and one bone image, which are bone suppression and bone enhancement, respectively. Moreover, the bone image is theoretically guaranteed to have better image contrast and the original input. We will prove this property in later section. 
\subsection{Mathematical Model}
We modify the dehazing model for natural images in Eq.~\ref{eq:model} to develop our model for X-ray images. First, we define the soft tissue image (background image) as
\begin{equation}
S(x,y)=A(1-t(x,y)),\,\mathrm{where}\, A=1\,.
\end{equation} Here, we assume $A=1$. The reason is that only X-ray can reach the sensors (there is no other light resource). We further define the unknown bone image $U(x,y)$ as a linear scaling of $J(x,y)$
\begin{equation}
U(x,y)=\frac{1}{\alpha}J(x,y)\,,
\end{equation} where $\alpha\ge 0$ is a scalar parameter. 
Taking these two equations into Eq.~\ref{eq:model}, we propose following model for X-ray images:
\begin{equation}\label{eq:ourmodel}
f(x,y)=\frac{1}{\alpha}U(x,y)(1-S(x,y))+S(x,y)\,,
\end{equation} where $f(x,y)\in [0,1]$ is the observed image, $U$ is the unknown bone image, $S(x,y)$ is the background image, $\alpha\ge 0$ is a scalar parameter. 

Our model keeps the physical meaning of the image intensity. When $f(x,y)=S(x,y)$, it would force $U(x,y)=0$. It means that the observation only comes from the background. When $f(x,y)=\frac{1}{\alpha}U(x,y)$, it would force $S(x,y)=0$. It indicates that the observation only comes from bones. Otherwise, the observation is composed by the background and bones as the similar way in natural images.

We use $\alpha$ as global constant variable, instead of spatially varying $\alpha(x,y)$. Although $\alpha(x,y)$ could achieve better visual result, it might introduce artifacts and it would lose the relationship between actual dose and image intensity in X-ray image. But when we use spatially constant $\alpha$, such linear scaling will keep such relationship between the actual physics and the intensity in X-ray images. 

In later section, we will prove that $\alpha\ge 1$ (Eq.~\ref{eq:alpha}), which theoretically guarantees to increase the image contrast. This property becomes clear when we set the background $B(x,y)=0$. That is $f(x,y)=\frac{1}{\alpha}U(x,y)$. It means $\nabla U(x,y)=\alpha \nabla f(x,y)$, where $\nabla$ is the standard gradient operator. Therefore, the contrast in bone image $U(x,y)$ is theoretically larger than the contrast in the input image $f(x,y)$. This theoretical property is numerically confirmed by all our experiments. 

In our model, if the background image $S(x,y)$ is already known, the bone image can be easily computed.
Therefore, we can solve our model by finding the $S(x,y)$ and $\alpha$. We first introduce some assumptions for our model. These assumptions will be used to estimate $S(x,y)$ in the following section. 

\subsection{Assumptions}
Since our model (Eq.~\ref{eq:ourmodel}) is ill-posed, we have to make some assumptions to solve this model. First of all, we assume $S(x,y)\le f(x,y)$. This assumption makes sure that $U(x,y)\ge 0$.  

Second, we assume $0\le S(x,y)<1$, which avoids the denominator to be zero. As a result, $\frac{1}{1-S(x,y)}>1$, which helps in improving the bone image contrast. Based on this assumption, we can prove $\alpha\ge 1$ in later sections (Eq.~\ref{eq:alpha}).

Third, background image $S(x,y)$ is smooth, especially for soft tissue region. More specifically, we assume $S(x,y)$ is second order differentiable~\cite{GONG2019329,gong:cf}. Such smoothness assumption is reasonable because the physical configuration of soft tissue is always smooth. As shown in following section, $S(x,y)$ can be obtained by solving a Laplace equation.

Fourth, we assume that the maximum value in $U(x,y)$ is one, to determine the value of $\alpha$. In later section, we can prove that $\alpha\ge 1$. Therefore, for most pixel locations (statistically), the contrast in the resulting bone image is always larger than the contrast in the original input image. In other words, the bone image is enhanced.

\subsection{Soft Tissue Image}
Now, we have enough assumptions to find $S(x,y)$. We estimate the background image by a two-step strategy. First, we roughly estimate a mask that covers bones. Be aware that the mask only needs to cover the bones. It does not necessarily align with bones' boundary. Therefore, there are multiple ways to generate such mask. It can be easily obtained by a simple threshold method followed by morphology operations. It can also be estimated by active contour methods. It can even be given interactively by users. In short, the way of obtaining this mask is flexible. Be aware that our mask only needs to cover the bones, but does not need to align the mask' boundary exactly on the bones' boundary. Thus, it is much easier than the bone segmentation task.

\begin{figure}[!tb]
	\centering
	\subfigure[Input Image]{\includegraphics[width=0.23\linewidth]{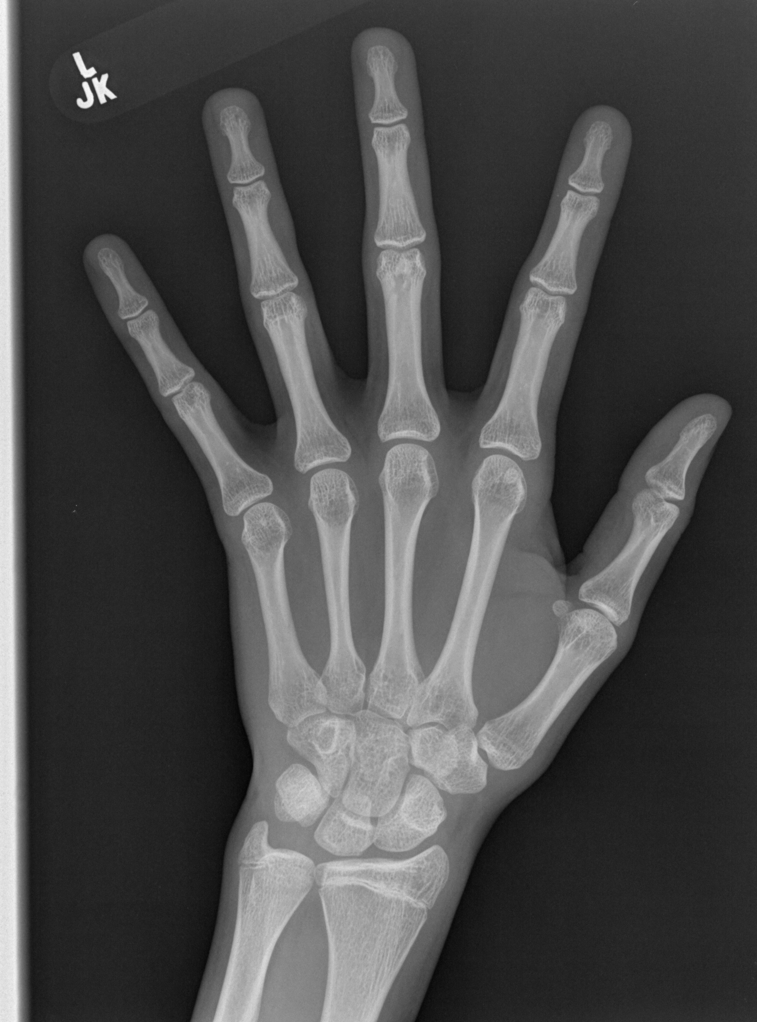}}
	\subfigure[Mask $M$]{\includegraphics[width=0.23\linewidth]{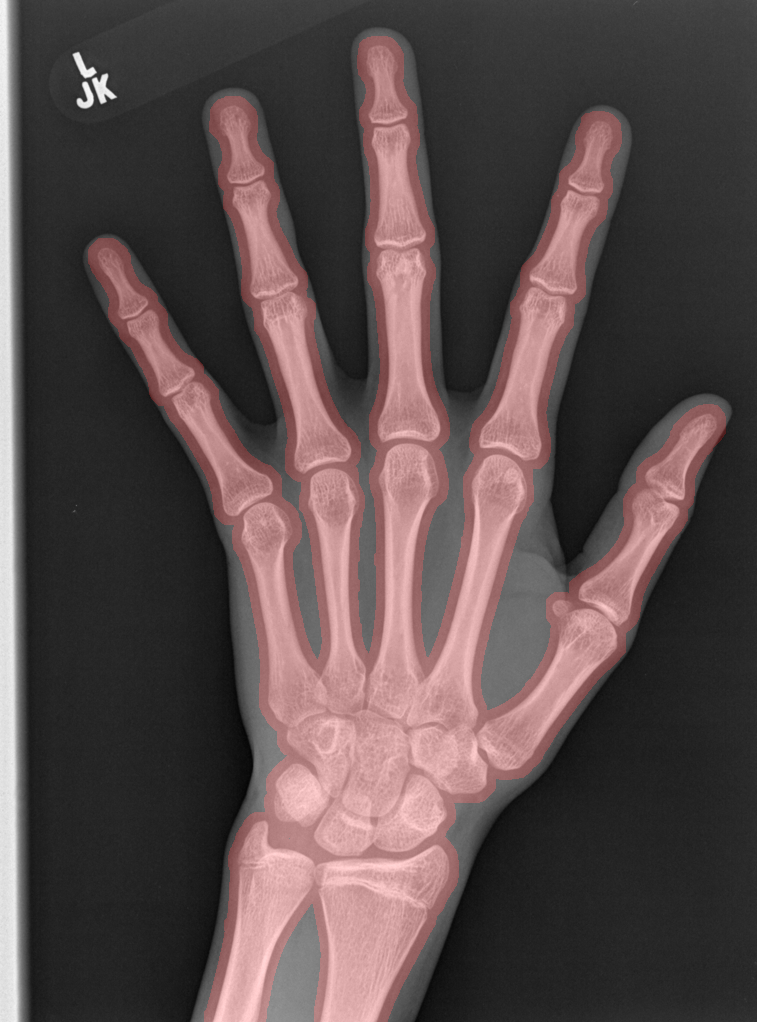}}
	\subfigure[Soft Tissue $S$]{\includegraphics[width=0.23\linewidth]{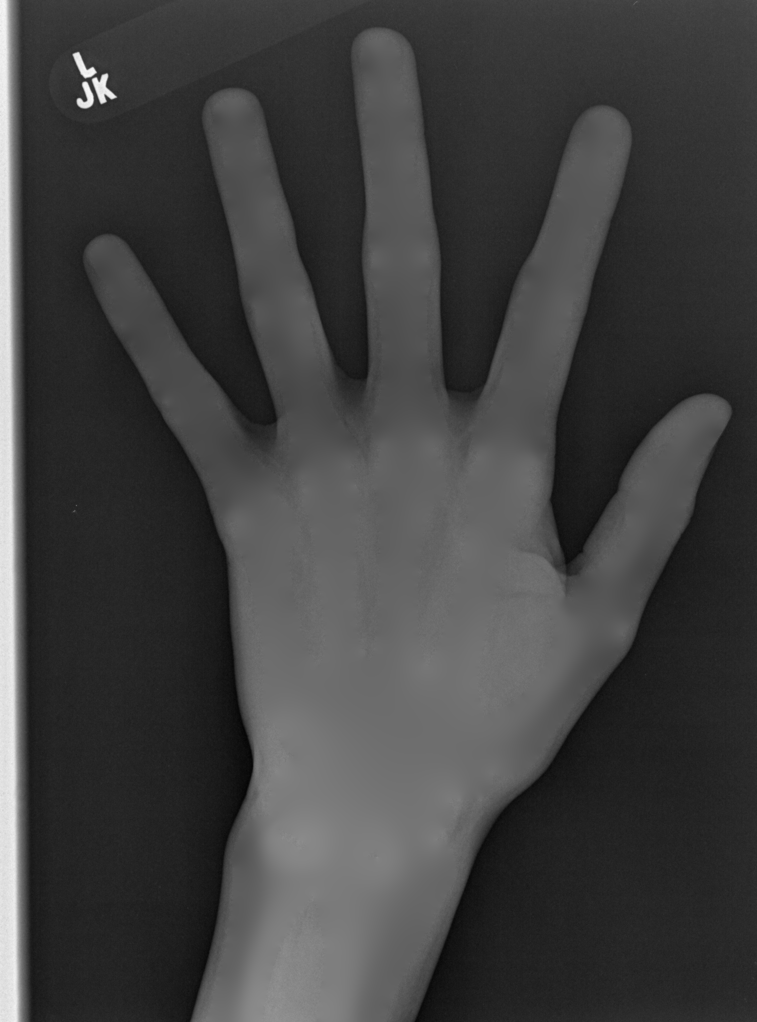}}
	\subfigure[Bones $U$]{\includegraphics[width=0.23\linewidth]{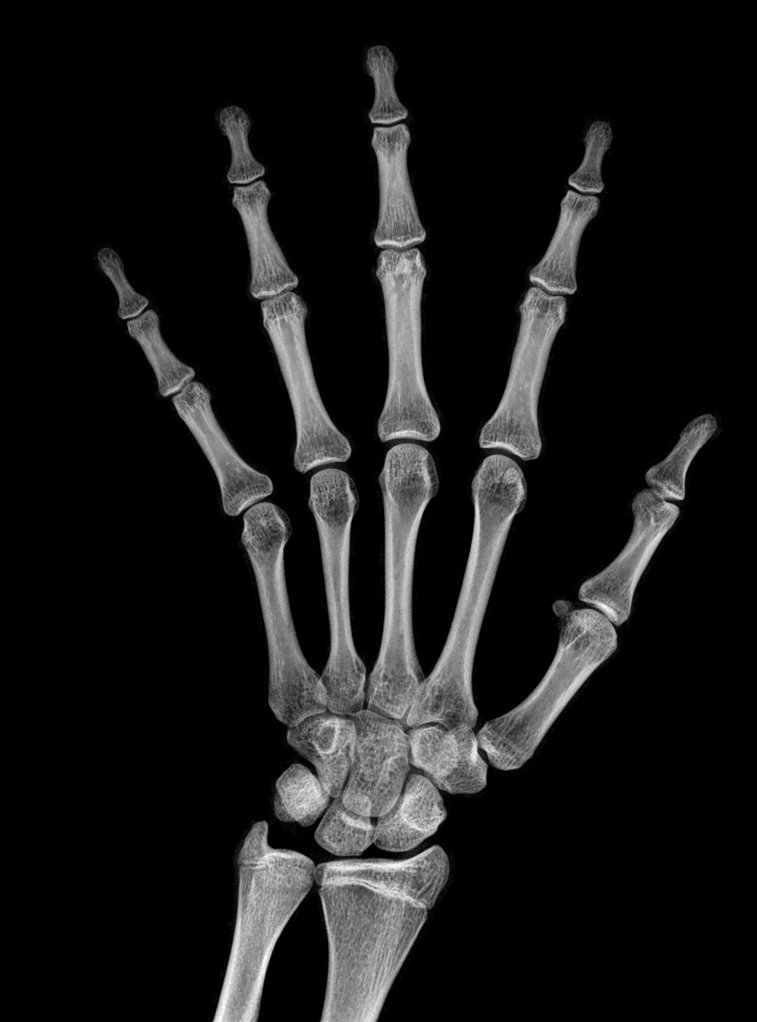}}
	\caption{From left to right: original X-ray image, our mask, estimated background and estimated bones. $S(x,y)$ is obtained by solving a Laplace equation. The bone image (d) has better contrast and the details on the bones become clear (enhanced).}
	\label{fig:exam}
\end{figure}

Second, we find $S(x,y)$ by solving a Laplace equation. Let $M(x,y)$ denote our mask. Now, we need to estimate the soft tissue intensity in this mask. This problem can be modeled as following minimization task 
\begin{equation}\label{eq:poisson}
\min \int\int_{M}||\nabla S||^2\,,\mathrm{s.t.}\,S_{\partial M}=f_{\partial M}\,,
\end{equation} where $\partial$ denotes the boundary. The optimal solution of this energy is the standard Laplace equation
\begin{equation}\label{eq:poisson2}
\Delta S_M=0\,,\mathrm{s.t.} \,S_{\partial M}=f_{\partial M}
\end{equation} This equation can be efficiently solved by the convolution pyramid method~\cite{Farbman2011}, which has linear computational complexity. The estimated background image is shown in Fig.~\ref{fig:exam}(c). 

Solving Eq.~\ref{eq:poisson2} is numerically efficient. The running time is 0.1 seconds in MATLAB on a ThinkPad P1 laptop with Intel Xeon E2176 CPU. The image resolution is $1022\times 757$. Such performance is enough for clinical applications in practice.

\subsection{Bone Image}
After estimating $S(x,y)$, we need to estimate $\alpha$ for bone image $U(x,y)$ estimation. As mentioned, we assume the maximum value in $U(x,y)$ is one. Therefore, we define
\begin{equation}\label{eq:alpha0}
\alpha\equiv\frac{1}{\max\{\frac{f(x,y)-S(x,y)}{1-S(x,y)}\}}\,.
\end{equation} Since $0\le f(x,y)\le 1$, we have $\frac{f(x,y)-S(x,y)}{1-S(x,y)}\le 1$. As a result, we can prove
\begin{equation}\label{eq:alpha}
\alpha\ge 1\,.
\end{equation}\.
This parameter linearly increases the contrast in the bone image. As mentioned, such linearity can keep the physical meaning of intensity in X-ray images.

\begin{figure}[!tb]
	\centering
	\subfigure[original]{
		\begin{overpic}[width=0.25\linewidth]{images/exam.png}
			\linethickness{1mm}
			\put(0,50){\color{blue}\line(100,0){75}}
	\end{overpic}	}
	\subfigure[middle line intensity profile]{\includegraphics[width=0.4\linewidth]{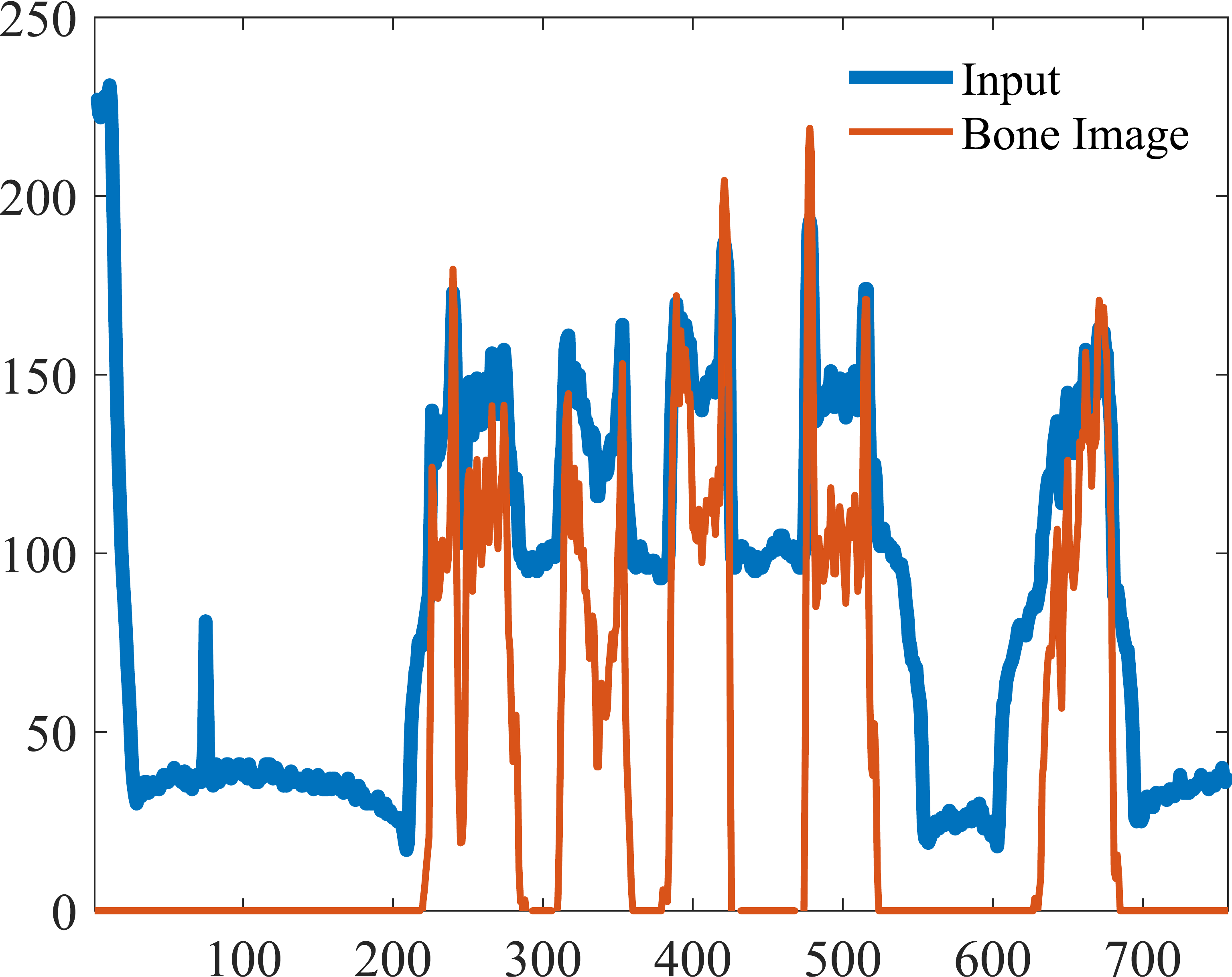}}
	\subfigure[our bone image]{
		\begin{overpic}[width=0.25\linewidth]{images/exam_U.png}
			\linethickness{1mm}
			\put(0,50){\color{red}\line(100,0){75}}
		\end{overpic}
	}

	\caption{From left to right: original images, the middle line intensity profile from the input (blue) and our result (red), our estimated bone image. In this case, $\alpha=1.44$ and the contrast (gradient) is increased, although the intensity might be lower.}
	\label{fig:examB}
\end{figure}

Finally, the bone image can be computed as (shown in Fig.~\ref{fig:exam}(d) and Fig.~\ref{fig:examB}(c))
\begin{equation}\label{eq:ourbone}
U(x,y)=\alpha\frac{f(x,y)-S(x,y)}{1-S(x,y)}\,.
\end{equation}

One example is shown in Fig~\ref{fig:exam}. And the middle line intensity profiles of original and our results are shown in Fig.~\ref{fig:examB}. In this case, $\alpha=1.44$ and the image contrast is enhanced. As shown in Eq.~\ref{eq:alpha}, $\alpha\ge 1$ and the enhancement is theoretically guaranteed. All our experiments also confirm this property.

Now, let us study the gradient of the resulting bone image to show the image contrast enhancement. From Eq.~\ref{eq:ourbone}, we can get the gradient of $U$
\begin{equation}
	\nabla U=\alpha[\frac{\nabla f}{1-S}-\frac{1-f}{(1-S)^2}\nabla S]\,.
\end{equation} Since we assume $S(x,y)$ is smooth, we know that $\nabla S(x,y)\approx 0$ for most locations~\cite{gong:gdp} (see the gradient statistics in Fig.5 from Ref.~\cite{GONG2019329}). Therefore, we have
\begin{equation}
\nabla U\approx\alpha\frac{\nabla f}{1-S}\ge \alpha\nabla f\ge \nabla f\,.
\end{equation} This result indicates that the bone image has better image contrast than the input image for most of pixels. Moreover, the larger $\alpha$, the better bone image contrast.

\subsection{Model Solver Summary}
In summary, our model in Eq.~\ref{eq:ourmodel} can be efficiently solved by Algorithm~\ref{algo:HW}.
\begin{algorithm}[!htb]
	\caption{Bone and Soft Tissue Decomposition}
	\label{algo:HW}
	\begin{algorithmic}
		\Require input X-ray image $f(x,y)$
		\State obtain the mask $M(x,y)$ by active contour or user input
		\State compute $S(x,y)$ by solving Eq.~\ref{eq:poisson2}
		\State compute $\alpha$ by Eq.~\ref{eq:alpha0}
		\State compute $U(x,y)$ by Eq.~\ref{eq:ourbone}
		\Ensure $S(x,y)$, $U(x,y)$
	\end{algorithmic}
\end{algorithm}

\section{Experiments}
We performed three experiments for our method. First, we perform our method on several X-ray images, showing our method is not restricted by specific imaging objects. Second, we compared our method with image enhancement method and dehazing method, showing that our modification of the original dehazing indeed helps in this task. Third, we perform our method on a hand X-ray image dataset, showing its effectiveness and efficiency.

\begin{figure}[!tb]
	\centering
	\subfigure[original]{\includegraphics[width=0.3\linewidth]{images/knees.png}}~~
	\subfigure[soft tissue]{\includegraphics[width=0.3\linewidth]{images/knees_B.png}}
	\subfigure[bone ($\alpha=1.34$)]{\includegraphics[width=0.3\linewidth]{images/knees_U.png}}
	\subfigure[original]{\includegraphics[width=0.3\linewidth]{images/humerus.png}}~~
	\subfigure[soft tissue]{\includegraphics[width=0.3\linewidth]{images/humerus_B.png}}
	\subfigure[bone  ($\alpha=1.08$)]{\includegraphics[width=0.3\linewidth]{images/humerus_U.png}}
	
	\subfigure[original]{\includegraphics[width=0.3\linewidth]{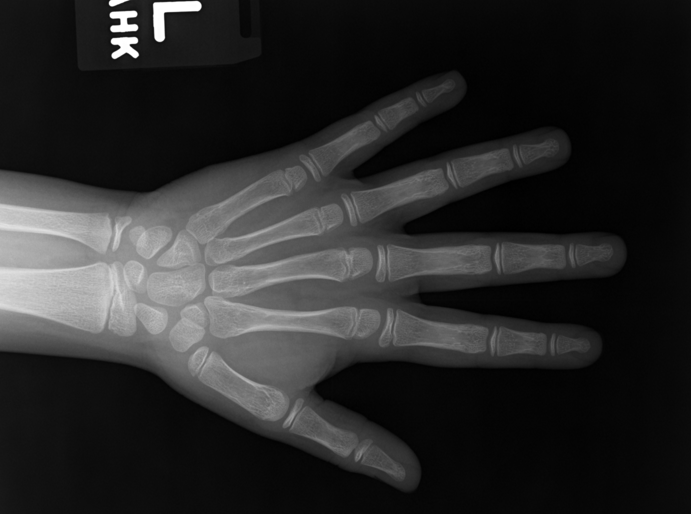}}~~
	\subfigure[soft tissue]{\includegraphics[width=0.3\linewidth]{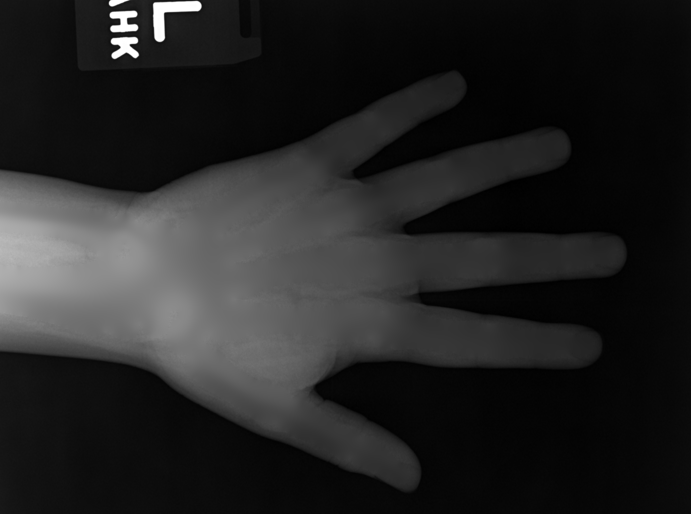}}
	\subfigure[bone ($\alpha=1.42$)]{\includegraphics[width=0.3\linewidth]{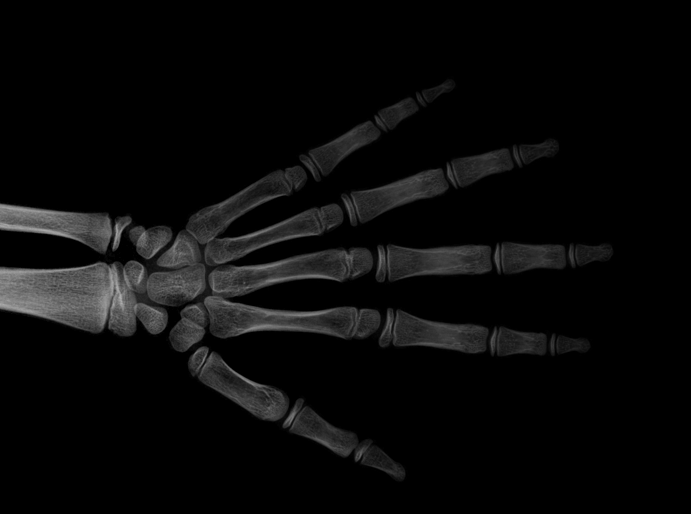}}
	
	\subfigure[original]{\includegraphics[width=0.3\linewidth]{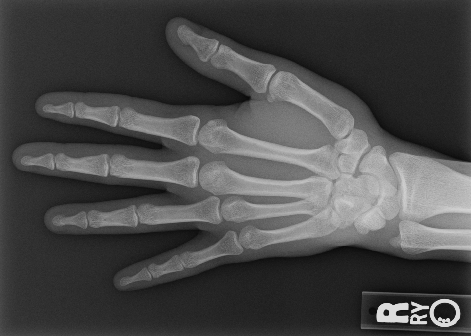}}~~
	\subfigure[soft tissue]{\includegraphics[width=0.3\linewidth]{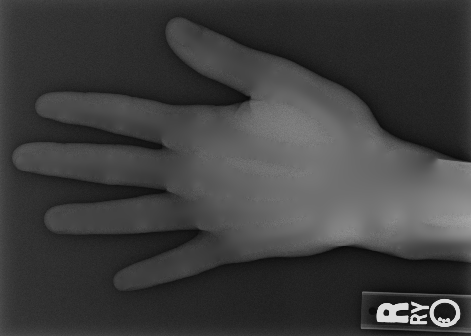}}
	\subfigure[bone ($\alpha=1.49$)]{\includegraphics[width=0.3\linewidth]{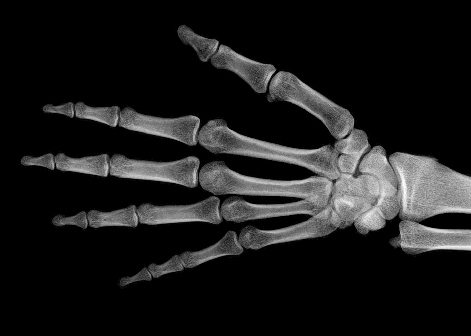}}
	\caption{More results by our method. Input X-ray images (left), our estimated soft tissue (middle) and estimated bone image (right).}
	\label{fig:ex}
\end{figure}

Several results from our method are shown in Fig.~\ref{fig:ex}. The left column is the original input image. The right two columns are the soft tissue and bone image, respectively. It can be told that the soft tissue image is smooth as we assumed. Meanwhile, the bone image has better image contrast as desired. Moreover, our method can reach real-time performance on these X-ray images. The running time of our method on these images is reported in Table~\ref{tab:1}.

\begin{table}
\centering
\caption{The running time in seconds of our algorithm}
\begin{tabular}{|c|c|c|}
\hline
image & resolution & time (seconds)\\
\hline
Fig.~\ref{fig:ex}(a) & 319$\times$442 & 0.031\\
\hline
Fig.~\ref{fig:ex}(d) & 193$\times$382 & 0.019\\
\hline
Fig.~\ref{fig:ex}(g) & 514$\times$711 & 0.094\\
\hline
Fig.~\ref{fig:ex}(j) & 336$\times$471 & 0.041\\
\hline
\end{tabular}
\label{tab:1}
\end{table}

\begin{figure}[!tb]
	\centering
	\subfigure[original]{\includegraphics[width=0.46\linewidth]{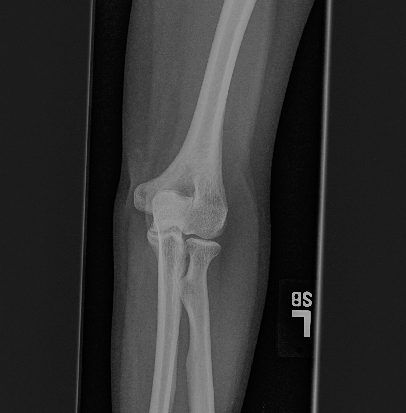}}
	\subfigure[histogram EQ]{
		\begin{overpic}[width=0.46\linewidth]{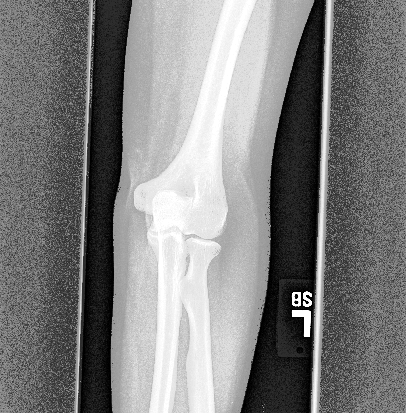}
			\put(10,60){\color{red}\vector(1,0){19}}
	\end{overpic}	}

	\subfigure[dehazing]{
		\begin{overpic}[width=0.46\linewidth]{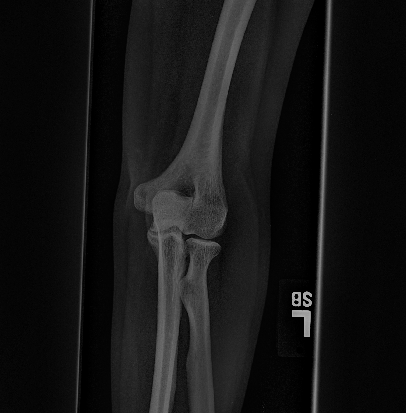}
			\put(10,60){\color{red}\vector(1,0){19}}
		\end{overpic}
	}
	\subfigure[ours $\alpha=2.08$]{
		\begin{overpic}[width=0.46\linewidth]{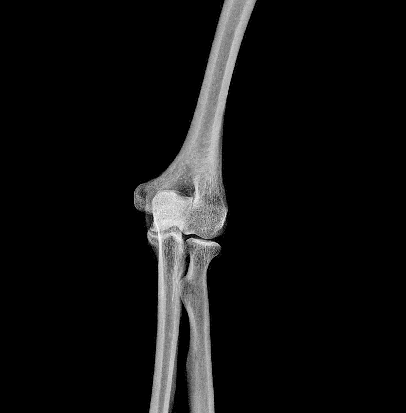}
			\put(10,60){\color{green}\vector(1,0){19}}
		\end{overpic}
	}
	
	\caption{(a) original images, (b) image enhancement by histogram equalization, (c) results from dehazing method with dark channel prior and guided filter, and (d) results from our method. The conventional methods can not completely remove the soft tissue (red arrows). Our method does not have this problem (green arrows).}
	\label{fig:exp1}
\end{figure}
\begin{figure}[!htb]
	\centering
	\subfigure[original]{\includegraphics[width=0.46\linewidth]{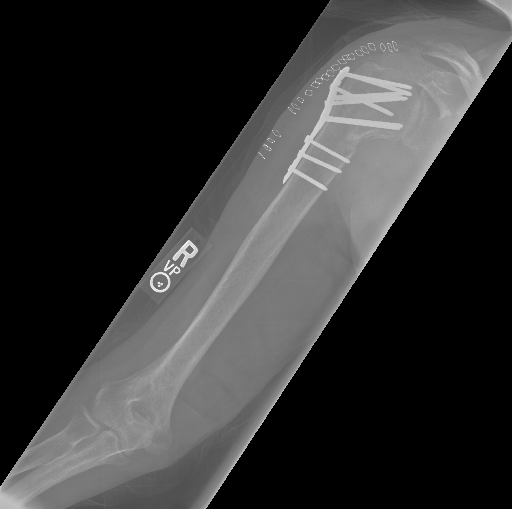}}
	\subfigure[histogram EQ]{
		\begin{overpic}[width=0.46\linewidth]{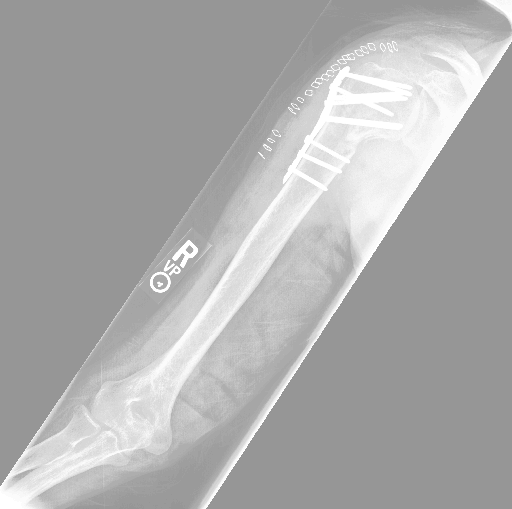}
			\put(48,55){\color{red}\vector(1,0){19}}
	\end{overpic}	}

	\subfigure[dehazing]{
		\begin{overpic}[width=0.46\linewidth]{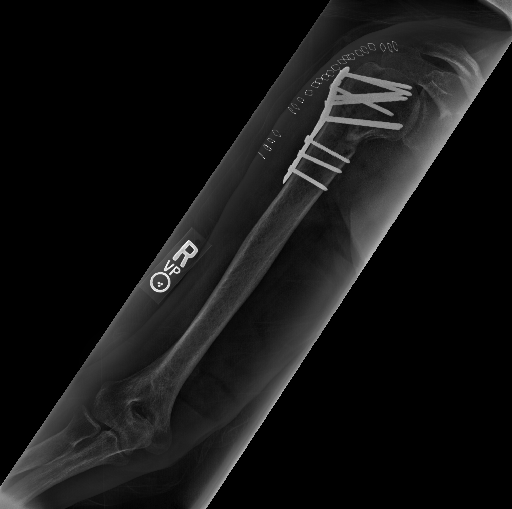}
			\put(48,55){\color{red}\vector(1,0){19}}
		\end{overpic}
	}
	\subfigure[ours $\alpha=1.43$]{
		\begin{overpic}[width=0.46\linewidth]{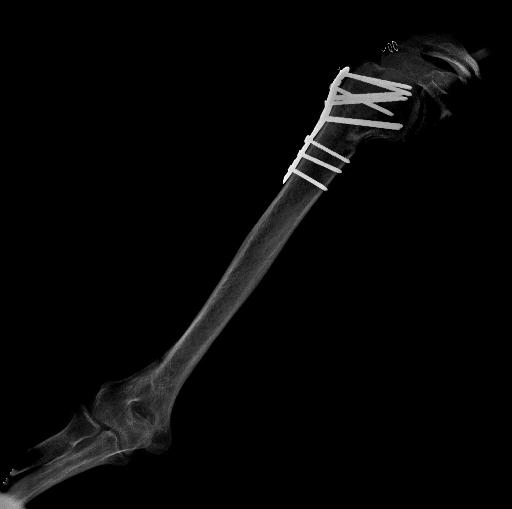}
			\put(48,55){\color{green}\vector(1,0){19}}
		\end{overpic}
	}
	
	\caption{(a) original images, (b) image enhancement by histogram equalization, (c) results from dehazing method with dark channel prior and guided filter, and (d) results from our method. The conventional methods can not completely remove the soft tissue (red arrows). Our method does not have this problem (green arrows).}
	\label{fig:exp2}
\end{figure}

We further compare our method with a classical image enhancement method and a dehazing method for natural images, which uses dark channel prior~\cite{DarkPrior}. The results are shown in Fig.~\ref{fig:exp1} and \ref{fig:exp2}. The classical image enhancement method (histogram equalization) enhances both the soft tissue and bones. And relationship between image intensity and physical X-ray is lost. 

Our model is different from the dehazing model. The dehazing method for natural images can not completely remove the soft tissue in X-ray image, as shown by the red arrows in Fig.~\ref{fig:exp1} and \ref{fig:exp2}. In contrast, our method does not have this issue. This is because we estimate a better soft tissue image. Moreover, our bone image has better image contrast, which is theoretically guaranteed as described in previous sections.
\begin{figure*}
	\centering
	\subfigure[resolution $2044\times1514$, run time 0.35s, $\alpha=1.34$ ]{\includegraphics[width=0.48\linewidth]{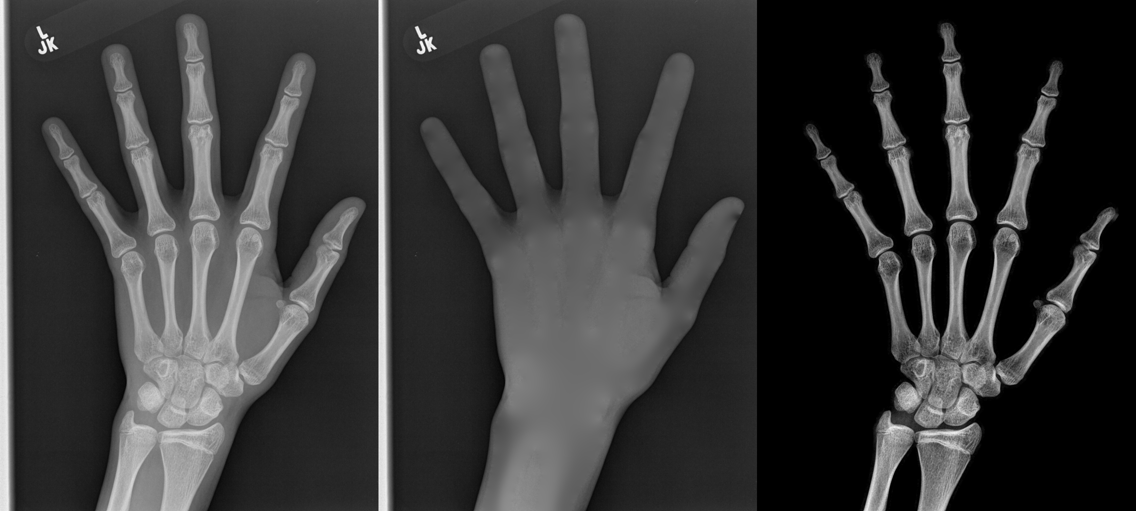}}~~\subfigure[resolution $2044\times1514$, run time 0.35s, $\alpha=1.39$ ]{\includegraphics[width=0.48\linewidth]{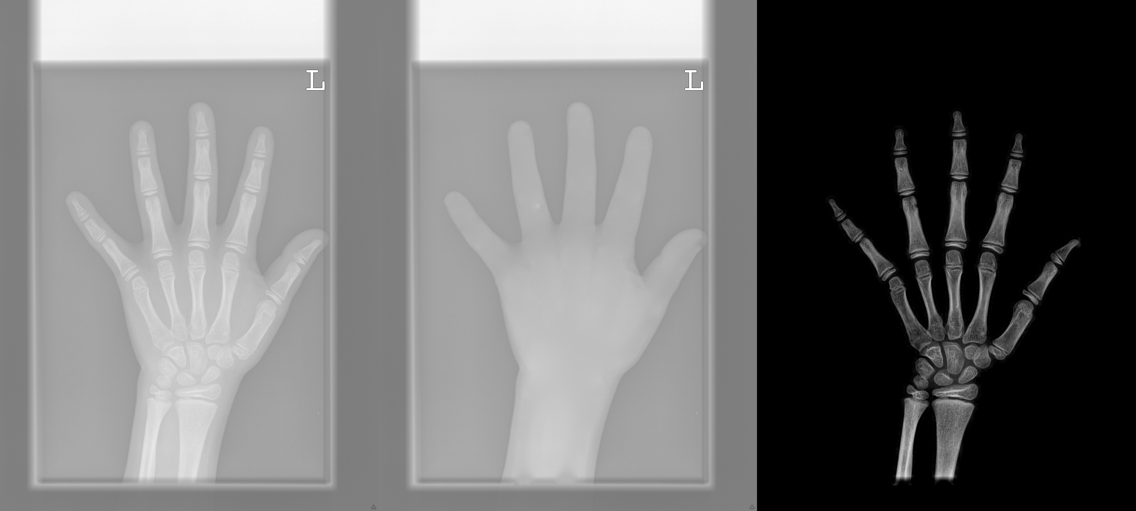}}
	
	\subfigure[resolution $2044\times1514$, run time 0.35s, $\alpha=1.63$ ]{\includegraphics[width=0.48\linewidth]{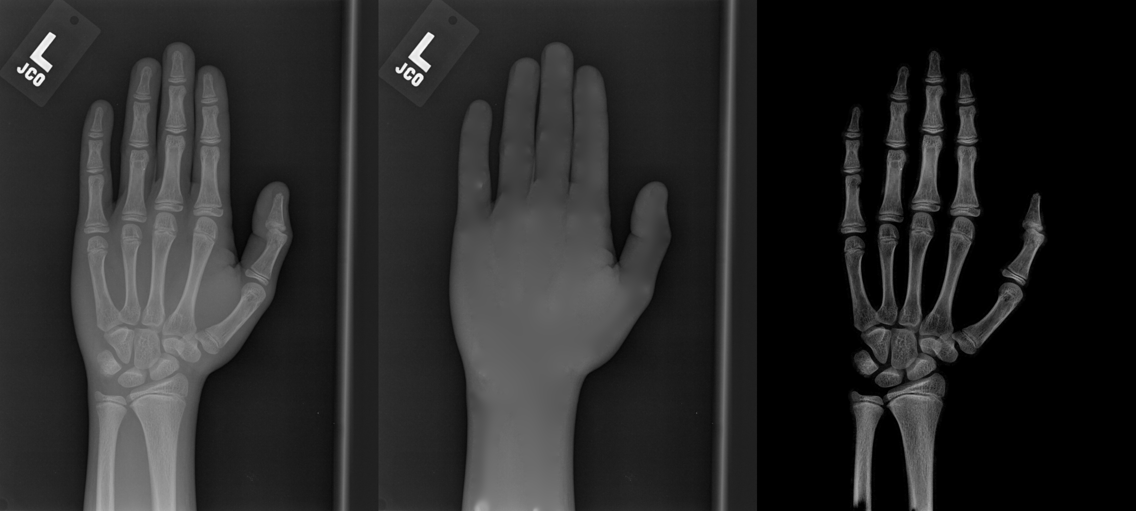}}~~\subfigure[resolution $2044\times1514$, run time 0.35s, $\alpha=2.42$ ]{\includegraphics[width=0.48\linewidth]{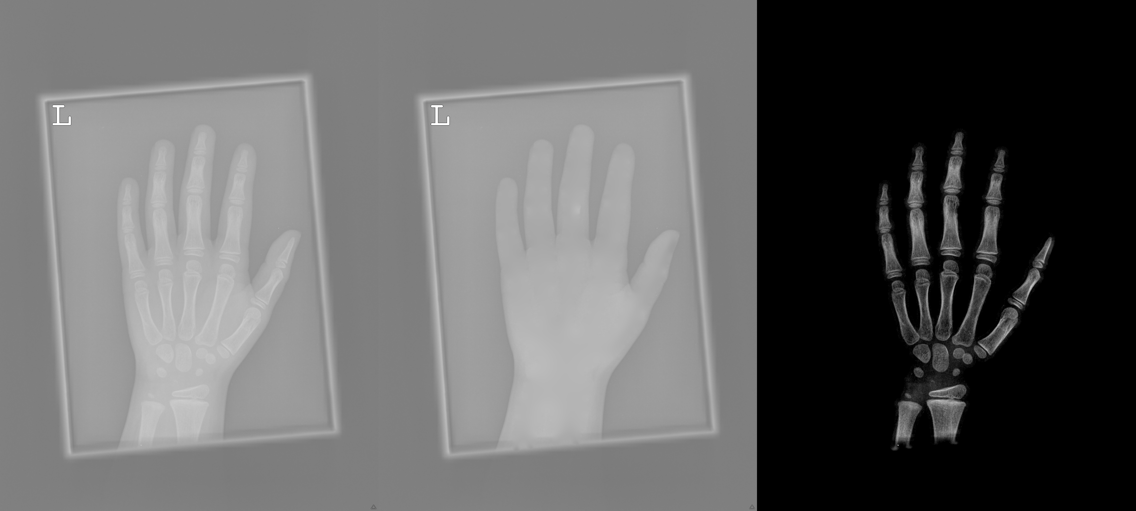}}
	
	\subfigure[resolution $2044\times1514$, run time 0.35s, $\alpha=2.31$ ]{\includegraphics[width=0.48\linewidth]{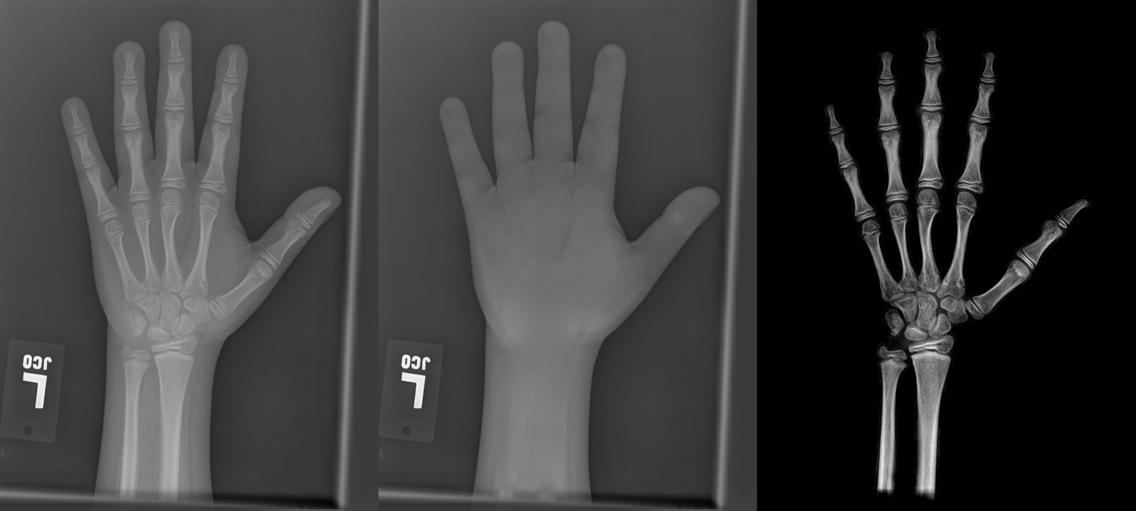}}~~\subfigure[resolution $2044\times1514$, run time 0.35s, $\alpha=3.61$ ]{\includegraphics[width=0.48\linewidth]{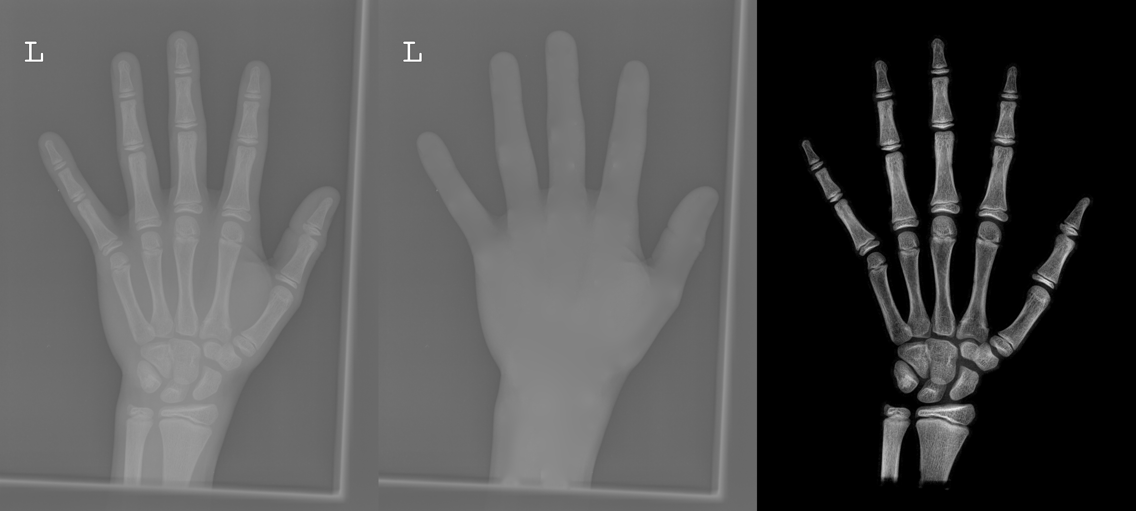}}
	
	\subfigure[resolution $2044\times1514$, run time 0.35s, $\alpha=3.02$ ]{\includegraphics[width=0.48\linewidth]{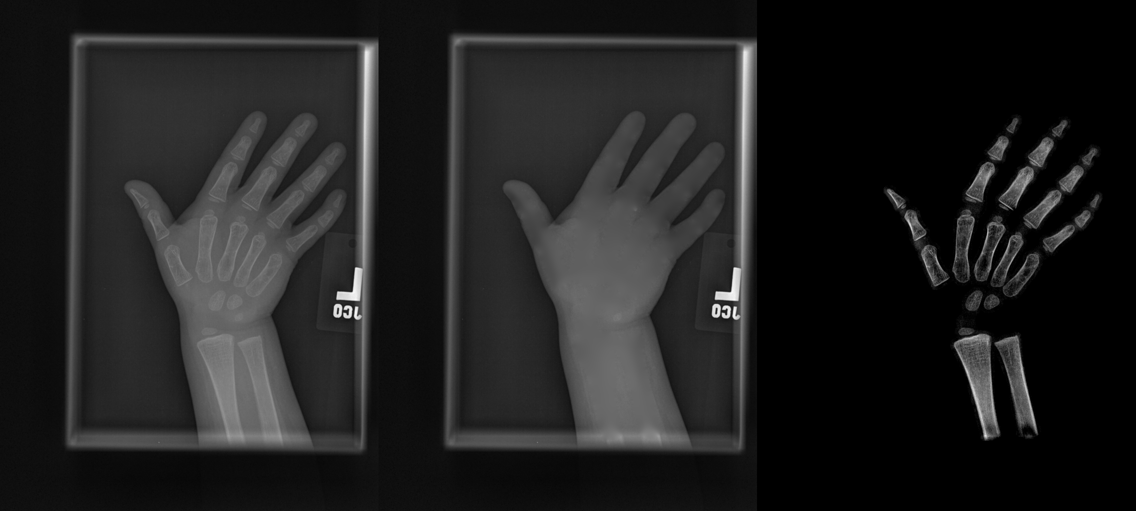}}~~\subfigure[resolution $2044\times1514$, run time 0.35s, $\alpha=1.47$ ]{\includegraphics[width=0.48\linewidth]{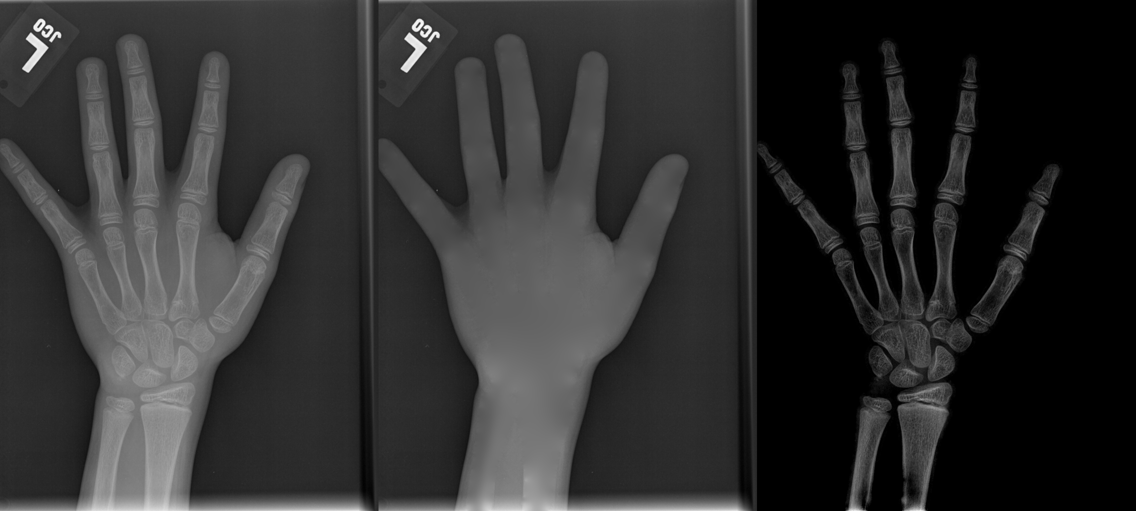}}
	
	\subfigure[resolution $2044\times1514$, run time 0.35s, $\alpha=2.11$ ]{\includegraphics[width=0.47\linewidth]{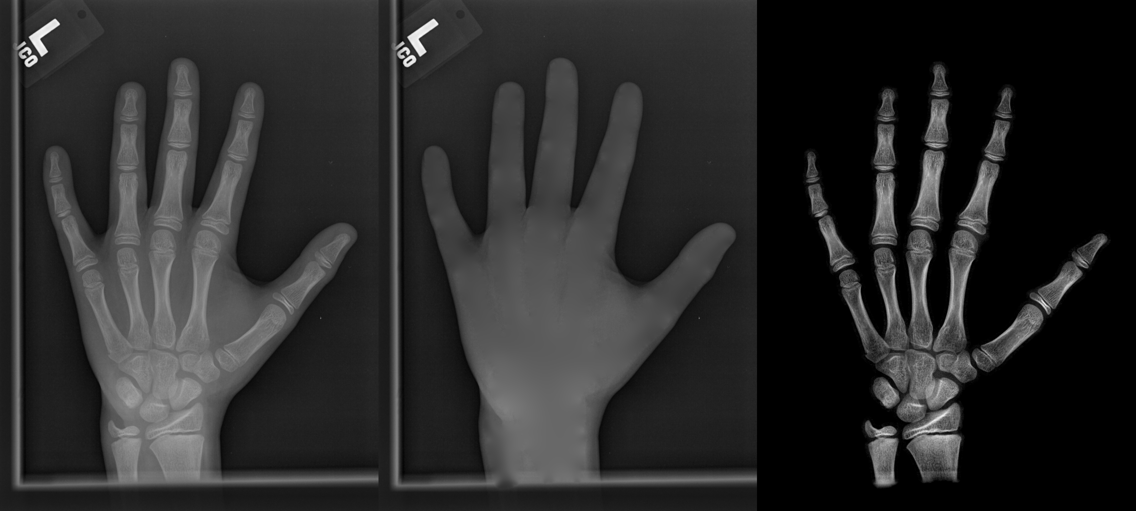}}~~\subfigure[resolution $2570\times2040$, run time 0.58s, $\alpha=2.34$ ]{\includegraphics[width=0.5\linewidth]{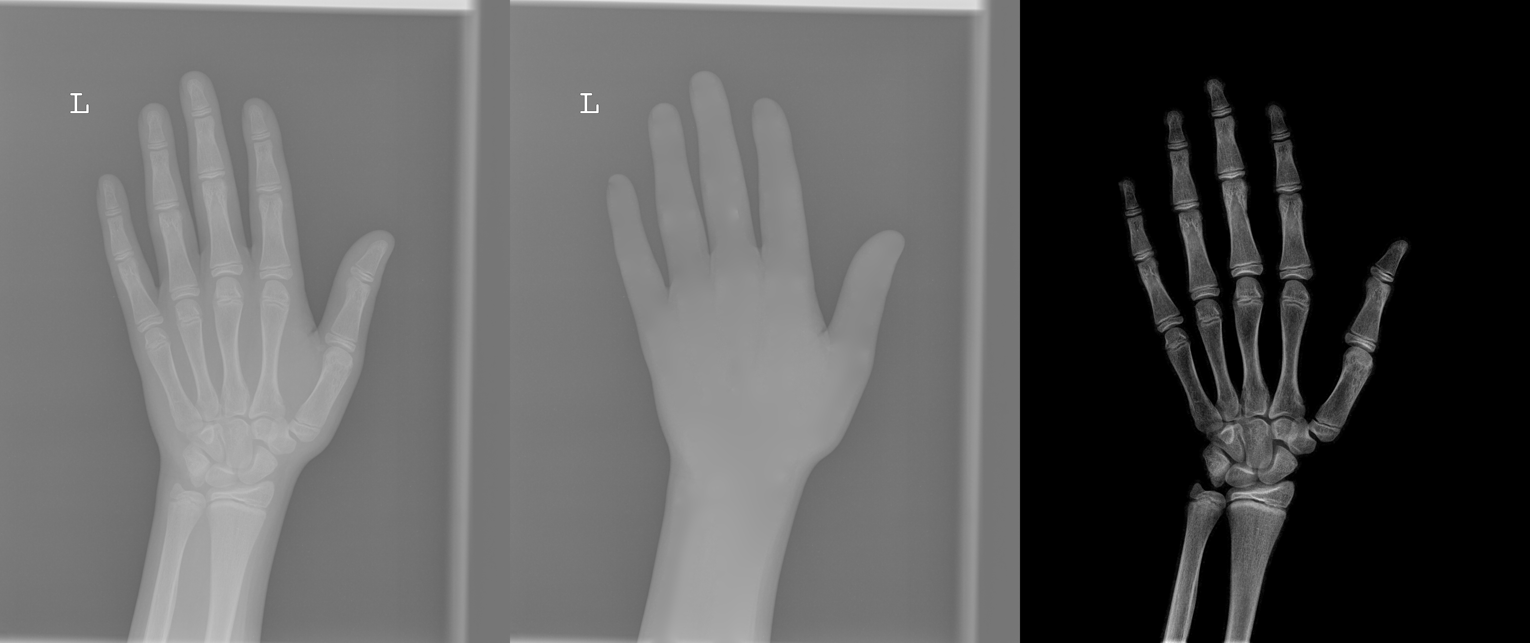}}
	\caption{From left to right in each panel: input X-ray images (left), our estimated soft tissue (middle) and estimated bone image (right). The resolution, running time of our algorithm and parameter $\alpha$ are provided. For these practical images, our method requires about half second to achieve the bone and soft tissue decomposition task in MATLAB language on a laptop. Higher performance can be achieved by C++ language on a better hardware.}
	\label{fig:last}
\end{figure*} 

In the third experiment, we applied our method on a hand X-ray image data set (RSNA), which contains more than 10,000 hand X-ray images. And the image has high resolution (usually larger than $1514\times2044$). These images are collected from clinical applications. Therefore, we can apply our method on these practical images, showing the efficiency and effectiveness of our method on real high resolution images.

In each panel of Fig.~\ref{fig:last}, the input image (left) is decomposed into soft tissue (middle) and bone image (right) by our method. Although we only show the first ten images from the data set, the results for the rest images are similar. 

The bone images have better image contrast since the parameter $\alpha\ge 1$ is theoretically guaranteed. Such enhancement can also be directly told by radiologists. Such enhancement is good for bone diagnosis in practical applications.   

Moreover, the running time of our method on such high resolution images is less than half second in the MATLAB language on a laptop. Therefore, it can achieve higher performance on a better hardware in real applications. If higher performance is required, our model can be solved by the parallel Laplace equation solver on a modern graphic process unit (GPU), which usually has thousands of cores. 

We believe that such bone and soft tissue decomposition is important for X-ray images, bone study, soft tissue diagnosis, etc. And the mathematical model can be very efficiently solved by solving a Laplace equation. 
\section{Conclusion}
In this paper, we propose to decompose one X-ray image into a soft tissue image and a bone image. We name this task as Bone and Soft Tissue Decomposition (BSTD). For this task, we develop a novel mathematical model. Our mathematical model is inspired by the natural dehazing model, but with proper extension for X-ray images. 

With several assumptions, our model leads to a Laplace equation, which can be efficiently solved. Solving the 2D Laplace equation is a classical problem. And we use the wavelet solver developed in~\cite{Farbman2011} to solve this equation. After solving this equation, we obtain the soft tissue image.

With the soft tissue image and the original input image, we can compute the scaling parameter $\alpha$. After getting the value of $\alpha$, we can compute the bone image with a close form solution expression. The bone image is uniquely determined by the soft tissue image. 

The resulting bone images are theoretically guaranteed to have better image contrast (larger gradient) because of $\alpha\ge 1$. Several numerical experiments have confirmed this property. Better image contrast is important for clinical diagnosis, such as bone fracture and surgery planning.

Our method can enhance the details on bones in X-ray images, without losing the relationship between the intensity and actual physical X-ray received on the sensor. This property is different from the conventional image enhancement methods. Our result can improve other bone related tasks, such as bone segmentation, recognition, diagnosis, surgery planning, etc. 

Moreover, our method is numerically fast. It can process 0.77 Million pixels per second in MATLAB software on a ThinkPad P1 laptop with Intel Xeon E2176 CPU. For real X-ray images with resolution $2044\times1514$, our method only requires 0.35 seconds to finish the bone and soft tissue decomposition task. 

Our method can be applied in a large range of applications. It can be used for bone study, for example, bone fracture diagnosis. It can also be used in bone age assessment, reducing the influence of soft tissue. Our method can also be used for applications where the soft tissue is the main concern, for example, pneumonia in chest X-ray images. Our method can be used as a pre-processing approach for deep learning training data set preparation.  
\if false
In the future, we plan to solve our mathematical model by modern convolution neural networks (CNN). Thanks to their excellent achievements in the past few years, CNN have been used in many different image processing and computer vision tasks. Our mathematical model is the loss function in such CNN. And our results from X-ray images can be used as training ground truth. We believe that the CNN can learn to generate the soft tissue and bone images from one input X-ray image. The network can be trained on the paired data $(f_i, S_i, U_i)$, where $f_i$ is the input image and $S_i$, $U_i$ are results from our method. 

\section*{Acknowledgment}
This work was supported in part by the National Natural Science Foundation of China under Grant 61907031.
\fi

%



%
\bibliographystyle{IEEEtran}
\bibliography{IEEEabrv,IP}

%




\end{document}